# Somatic mutations in human ageing: New insights from DNA sequencing and inherited mutations

Kasit Chatsirisupachai[1,2], João Pedro de Magalhães[1,3]

[1] Integrative Genomics of Ageing Group, Institute of Ageing and Chronic Disease, University of Liverpool, Liverpool L7 8TX, UK

[2] Current address: European Molecular Biology Laboratory (EMBL), Genome Biology Unit, Heidelberg, Germany

[3] Current address: Institute of Inflammation and Ageing, University of Birmingham, Queen Elizabeth Hospital, Mindelsohn Way, Birmingham, UK

**Abstract**

The accumulation of somatic mutations is a driver of cancer and has long been associated with ageing. Due to limitations in quantifying mutation burden with age in non-cancerous tissues, the impact of somatic mutations in other ageing phenotypes is unclear. Recent advances in DNA sequencing technologies have allowed the large-scale quantification of somatic mutations in ageing. These studies have revealed a gradual accumulation of mutations in most normal tissues with age as well as a substantial clonal expansion driven mostly by cancer-related mutations. Nevertheless, because of the relatively modest burden of age-related somatic mutations identified so far and their stochastic nature, it is difficult to envision how somatic mutation accumulation alone can explain most ageing phenotypes that develop gradually. Studies across species have also found that longer-lived species have lower somatic mutation rates, though these could be explained by selective pressures to reduce or postpone cancer as longevity increases. Overall, with a few exceptions like cancer, results from recent DNA sequencing studies do not add weight to the idea that somatic mutations with age drive ageing phenotypes and the phenotypic role, if any, of somatic mutations in ageing remains unclear. Recent studies in patients with somatic mutation burden and no signs of accelerated ageing further question the role of somatic mutations in ageing.

**Introduction**

As the blueprint to life and cellular functions, one of the primary suspects as a driver of ageing is the gradual accumulation of DNA damage and mutations. Studies in rodents suggest that disruption in DNA repair and DNA damage responses can often result in a short lifespan and phenotypes resembling accelerated ageing, known as 'segmental progeroid syndromes' (Freitas & de Magalhaes 2011; Vijg 2021). Human progeroid syndromes, like Werner Syndrome, also originate from mutations in genes involved in DNA damage responses. Evaluating and quantifying DNA damage and mutations in vivo is not straightforward, however, not least because of their random nature. In rodents, gene reporter systems have been used to estimate levels of DNA damage and somatic mutations, showing a gradual increase in most (but not all) tissues (Freitas & de Magalhaes 2011; Ren *et al.* 2022). Given the age-related incidence of cancer and the higher number of mutations in tumours from older patients (Chatsirisupachai *et al.* 2021), mutations are expected to accumulate with age in human tissues, but until recently with limited empirical evidence.

Recent advances in high throughput technologies and particularly DNA sequencing provide new opportunities to investigate and gain insights into ageing and other diseases and processes (de Magalhaes *et al.* 2010; Ren *et al.* 2022). Studies during the past few years have revealed the landscape of somatic mutations, particularly single-nucleotide variants (SNVs), in non-cancerous tissues, which we outlined below. Recent comparisons in mutation rates across species are also discussed. While somatic mutation burden increases with age in all tissues investigated, the relevance of the mutation accumulation to ageing processes (other than cancer) remains unclear. Due to the modest number of mutations detected, we argue that they are unlikely to explain most ageing phenotypes.

**Somatic mutations accumulate with age in non-cancerous tissues**

Several recent studies using bulk genome and transcriptome sequencing have identified somatic mutations in human tissues, including strong age-related patterns (Blokzijl *et al.* 2016; Lodato *et al.* 2018; Martincorena 2019; Yizhak *et al.* 2019; Abascal *et al.* 2021; Li *et al.* 2021; Moore *et al.* 2021; Ren *et al.* 2022; Rockweiler *et al.* 2023). Unlike in tumours, mutations in normal tissues typically occurred in a subset of cells within the tissue. Thus, these studies usually performed very high-depth bulk sequencing on either single-cell-expanded clones (Blokzijl *et al.* 2016) or on laser-captured clonal structures (Lee-Six *et al.* 2019). Clearly, there is an increase with age in somatic mutations (Evans & DeGregori 2021; Marongiu & DeGregori 2022; Ren *et al.* 2022), as observed in various dividing and non-dividing tissues like the brain (Lodato *et al.* 2018), blood (Genovese *et al.* 2014; Jaiswal *et al.* 2014), colon (Lee-Six *et al.* 2019), heart (Choudhury *et al.* 2022), oesophagus (Martincorena *et al.* 2018; Yokoyama *et al.* 2019), skin (Martincorena *et al.* 2015), lung (Yoshida *et al.* 2020), testes (Maher *et al.* 2018), small intestine (Wang *et al.* 2023a), liver (Brunner *et al.* 2019) and others. In addition, two recent studies examined somatic mutations across several cell types/organs within the same individuals (Li et al Nature 2021, Moore et al Nature 2021). While numbers of SNVs detected in most tissues were found in a range of a few hundred to several thousand per genome, these numbers vary from tissue to tissue. For example, the highest mutation rate among tissues was found in the intestine and in colonic crypts, with about 50 base substitutions per cell per year, making up a total burden at approximately 5,000 SNVs of an 80-year-old individual. Conversely, the mutation rate was lowest in spermatogonia (Moore et al Nature 2021). Overall, across multiple studies of mutation rates in human tissues with age, SNVs/cell/year varied from ~2.4 in the testis to ~56 in the intestine (Ren *et al.* 2022).

Because somatic mutations occur largely randomly and their low abundance in the genome, bulk sequencing is only able to detect clonally amplified mutations, precluding the quantification of mutations occurring in only one or few cells. Recent studies thus developed single-cell approaches to detect somatic mutations in B lymphocytes (Zhang *et al.* 2019), hepatocytes (Brazhnik *et al.* 2020), bronchial epithelial cells (Huang *et al.* 2022), neurons (Miller *et al.* 2022), and cardiomyocytes (Choudhury *et al.* 2022). Not surprisingly, these studies reveal mutation accumulation rates during ageing consistent with those estimated from bulk sequencing. These rates vary among cell types. For example, somatic mutations accumulated ~3 times faster in cardiomyocytes than in neurons (Choudhury *et al.* 2022).

Mutational signature analysis revealed similarities and differences between the underlying mechanisms across cell types. For instance, age-related increases in the clock-like C>T mutational signature and in a signature associated with oxidative damage are observed in cardiomyocytes, neurons, lymphocytes and hepatocytes. However, only in cardiomyocytes the mismatch repair defect signature increases in an age-dependent manner (Choudhury *et al.* 2022). Interestingly, one study showed that age-related accumulation of somatic mutations in normal neurons is primarily due to the clock-like mutational process, while neurons from Alzheimer's disease patients harboured additional C>A variants associated with oxidative damage signature (Miller *et al.* 2022). Somatic mutations in Alzheimer's disease neurons were distributed across the genome and did not overlap with Alzheimer's disease risk loci, suggesting that these mutations were more likely to be a secondary effect caused by a high accumulation of ROS in neurons from Alzheimer's disease patients (Miller *et al.* 2022). Overall, both bulk and single-cell sequencing studies observed an accumulation of somatic mutations with age across human tissues.

The observation that mutations accumulate with age in non-dividing cells such as neurons argue against the idea of cell division as the key source of somatic mutations (Abascal *et al.* 2021). Strikingly, recent studies in human tissues indicate that the rate of somatic mutation accumulation with age does not appear to accelerate at older ages, as endogenous clock-like mutational processes, involving specific mutational signatures, accumulate linearly with age and account for most somatic mutations in normal tissues. Quite the contrary, the mutation rate is relatively high during the first embryonic divisions before dropping considerably, which could relate to the activation timing of more mature DNA repair mechanisms (Coorens *et al.* 2021). One recent analysis of somatic mutations in human brains found hypermutable brains were more frequent in individuals over 60 years-old, but when excluding hypermutable samples found no correlation of mutation burden with age, suggesting an early developmental origin of somatic mutations (Bae *et al.* 2022).

Taken together these recent results clearly confirm the idea that mutations accumulate with age in at least some human tissues. There is a very large variation between cells in the mutations detected, and all tissues are mosaics made up of genetically diverse individual cells, even though most mutations will be neutral (Mustjoki & Young 2021). This is unsurprising because mutations are sporadic and given the size of the genome mutations will occur in different genomic regions in different cells. If it happens for a mutation to result in increased cell proliferation or somehow improve the fitness of the cell (e.g. mutations in cancer driver genes), then this will result in the expansion of that clone (Martincorena 2019). Somatic mutations could then predispose to cancer, in line with the multi-hit hypothesis driving cancer. Data from the oesophagus, for instance, supports this concept of clonal expansion (Martincorena *et al.* 2018; Yokoyama *et al.* 2019). The degree of clonal expansion depends heavily on the tissue cell dynamics (e.g., whether the tissue self-renews or not) and anatomy. In some tissues, like the oesophagus, clonal expansion can result in large parts of an aged tissue being composed of mutant cell populations. In the colon, cells are organized into crypts, with each crypt containing ~2000 cells originating from stem cells at the base of a crypt. Thus, a clonally expanded clone usually localises within a crypt, highlighting an impact of tissue anatomy on clonal expansion. Although it may predispose to cancer, the impact clonal expansion has on normal physiology is not clear.

In some cases of stress and pathology, clonal expansions and mutations have been observed, usually predisposing to cancer, though clonal expansions can be protective as well (Kakiuchi & Ogawa 2021; Ogawa *et al.* 2022). Liver stress, such as chronic liver disease, can cause mutations in metabolism genes in liver that lead to clonal expansion (Ng *et al.* 2021), some of which are adaptive and protective (Zhu *et al.* 2019; Wang *et al.* 2023b). Mutations and clonal expansion may also contribute to inflammatory bowel disease (Olafsson & Anderson 2021). Likewise, some stressors can result in a significantly higher somatic mutation accumulation. Of note, mutation accumulation is substantially higher in the lungs of smokers, and higher in smokers than what is observed in aged non-smokers (Yoshida *et al.* 2020; Huang *et al.* 2022). One study found that smoking adds 1,000 to >10,000 mutations per cell (Yoshida *et al.* 2020). More specifically, although single base substitutions accumulate with age in the lung, at an estimated rate of 22 SNVs/cell/year, ex-smokers have an average increased burden of 2330 SNVs per cell and current smokers 5300 more SNVs per cell (Yoshida *et al.* 2020), which explains the much higher cancer incidence in smokers.

It should also be noted that these studies were representative of mutation landscape based on less than a hundred genomes from only a few individuals. Furthermore, the difficulties to detect somatic copy-number alteration (SCNAs) and structural variations (SVs, which include insertions and deletions) mean that current methods may be underestimating the actual number of alterations. Only 2-3 per cent of the elderly contains detectable clonal mosaicism from chromosomal anomalies in blood (Jacobs *et al.* 2012; Laurie *et al.* 2012). Recent studies also reported that SCNAs were rarely detected in non-cancerous solid tissues including oesophagus (Yokoyama *et al.* 2019; Li *et al.* 2021), liver (Brunner *et al.* 2019), colon (Lee-Six *et al.* 2019), and endometrium (Moore *et al.* 2020). However, it is expected that the continuous development of sequencing technology will allow a more precise detection of large-scale somatic variants. Indeed, more studies are needed to paint a more complete picture on somatic mutations in normal tissues, in particular regarding other types of mutations than SNVs.

**Mutation rates, cancer, and ageing across species**
Across species, one recent study using cell lines reported that mutagen-induced mutation frequencies inversely correlated with species-specific maximum life (Zhang *et al.* 2021). Another recent work from Cagan et al. also reported gradual increases in somatic mutations with age in intestinal crypts in various mammalian species, with a rate of somatic mutation accumulation inversely correlating with the lifespan of the species (Cagan *et al.* 2021). While this result could be interpreted as supporting the somatic mutation theory of ageing, it could also be argued that the strong negative association between somatic mutation rate and lifespan could come from increased cancer incidence rate in species with a higher rate of mutation accumulation. Indeed, cancer is widespread amongst metazoans (Albuquerque *et al.* 2018), occurring even in young animals, and selection to avoid or postpone cancer must be very strong in mammals, as further discussed below.

**Impact of age-related somatic mutation accumulation on human ageing**

Recent results showing an increase in somatic mutations with age have been interpreted by some as supporting the idea that mutation accumulation is important in ageing (Vijg & Dong 2020). On the other hand, it can be argued, as many have before (Maynard Smith 1959; de Grey 2007), that the number of mutations observed are too modest to envision widespread detrimental effects in tissues. Mutations are nearly always heterozygous and most of them affect non-coding regions, with only about 12 mutations in pancreas parenchyma to 76 mutations in liver exomes of the older ages (85 – 93 years old) (Li et al Nature 2021). In fact, mutations in non-coding gene regulatory regions are low as well. For example, a study found an average of 2.25 and 11.5 mutations per genome in promoters and enhancers of old satellite cells (Franco *et al.* 2018). Besides, patterns of selection in cancer and somatic tissues show that homozygous mutations are much more likely to be detrimental (Martincorena *et al.* 2017), and the probability of homozygous mutations in the absence of clonal expansion are extremely low – and nearly zero for homozygous mutations in the same gene in two nearby cells. While cancer driver mutations can lead to clonal expansion in tissues and ultimately cancer, no empirical evidence supports similar processes in ageing.

Possible mechanisms for how somatic mutations could drive ageing have been proposed, such as mutations in lymphocytes leading to pathogenic autoantibodies that contribute to autoimmune diseases (Singh *et al.* 2020), and through positive selection on mutations that lead to clonal expansions of phenotypically aberrant cells (Cagan *et al.* 2022). It has been suggested that one possible mechanism somatic mutations could contribute to ageing is through introducing gene expression noise (Vijg 2021). Somatic mutations that occur in regulatory regions of the genome could cause dysregulation in the gene regulatory network and aberrant transcription, leading to the physiological decline of tissue functions. Indeed, cell-to-cell variability in gene expression increases with age and correlates with an age-related increase in somatic mutations (Enge *et al.* 2017; Levy *et al.*

2020). Nonetheless, it is unclear how the stochastic nature of mutations and gene expression noise might result in gradual changes during ageing. Even if somatic mutations could contribute to gene expression noise, quantitative analysis is needed to demonstrate how many mutations per genome are required to cause enough noise that impacts cellular phenotypes; and whether this number of mutations is consistent with the limited number of mutations accumulated during ageing. Furthermore, the proportion of cells with altered mutation-created transcriptional noise in a tissue sufficient to disrupt tissue functions is currently unknown. Overall, the phenotypic consequences of mutations in ageing tissues remains unknown.

Ageing, and contrary to cancer, is not a result of the increase of cell proliferation. Quite the opposite, ageing and cancer could be seen as two sides of the same coin in that while cancer increases cell proliferation, ageing often results in a reduced proliferation typically accompanied by a loss of function, cells, and tissue mass. Recent results support this observation and have revealed that ageing and cancer have opposite transcriptional responses in most human tissues (Chatsirisupachai *et al.* 2019). In other words, cancer can be seen as an excess of cell proliferation and function while ageing is a loss of function often accompanied by loss of cells and tissue mass. That said, other ageing diseases besides cancer exhibit stochastic patterns, like clonal haematopoiesis of indeterminate potential, in turn a risk factor for cancer and cardiovascular disease (Jaiswal *et al.* 2017). Mutations may also drive specific age-related diseases (Mustjoki & Young 2021), like cerebral cavernous malformations that grow through a three-hit cancer-like mechanism (Ren *et al.* 2021). Clearly, some ageing phenotypes and diseases are likely driven by mutations. Such cases, other than cancer, appear to be a small subset of ageing phenotypes that are random and of low incidence overall. It does not seem to apply to most ageing phenotypes that are gradual and/or widespread (in most cases inevitable) like sarcopenia, loss of wound healing with age, menopause, ageing of the cardiovascular system, loss of sensory function, hair greying, loss of kidney function, cognitive ageing, thymus involution, loss of lung capacity, etc.

If random events like mutations are driving gradual ageing changes like sarcopenia, cognitive ageing and hair greying then one would expect these phenotypes to be random, to affect different parts of the tissue and/or different parts of the body at different times in different places. But by and large that is not what is observed. With exceptions like the above-mentioned cancer and cavernomas, ageing is fairly regular and consistent. As people age there are gradual changes, such as loss of muscle mass, hair greying and baldness, as well as a gradual loss of function in many organs (e.g. heart, lungs, and kidneys). Although there is variation in ageing patterns between individuals, they do not tend to vary as much within individuals. As an example, hair greying in the beard of men tends to be symmetrical (Poljsak *et al.* 2020). Therefore, we speculate that stochastic events like mutations are unlikely to be causal for most ageing phenotypes, in particular in light of their low abundance as recently revealed by next-generation sequencing.

**Causal evidence from inherited mutations in mice and men**

Apart from cancer and a small subset of diseases like cerebral cavernous malformations, causal evidence for a role of somatic mutations in ageing and age-related diseases is lacking (Olafsson & Anderson 2021). One recent study sequenced the genome of centenarians and revealed genetic variants in genes related to DNA repair as one possible mechanism (Garagnani *et al.* 2021), but correlation does not imply causation. It is therefore important to consider the causal evidence of a role of somatic mutations in ageing from specific genetic manipulations in mice and inherited diseases in humans.

Progeroid syndromes, most – but not all (Dolle *et al.* 2006) – of which in humans and mice are caused by defects in genome maintenance, support the idea that DNA damage is important in ageing (Vijg 2021; Franco *et al.* 2022). Nonetheless, some authors have questioned whether progeroid syndromes reflect normal ageing processes (Keshavarz *et al.* 2023), and the exact molecular mechanisms and specifically a role of somatic mutations in progeroid syndromes remains unclear. Besides, if somatic mutations drive ageing, then one would expect other defects resulting in a higher burden of somatic mutations to result in accelerated ageing. But that is not the case in mice or in humans. Several mouse models with elevated mutations fail to exhibit, apart from enhanced cancer, a clear acceleration of aging phenotypes (Kennedy *et al.* 2012; Franco *et al.* 2022). Examples include *Pms2*-null mice with a 100-fold higher mutation frequency in multiple tissues (Narayanan *et al.* 1997) and mutator mice with defects in DNA polymerase that have a mutation rate 17 times that of wild-type mice (Uchimura *et al.* 2015) and do not manifest progeroid symptoms, although they have a higher incidence of cancer and a shorter lifespan (Goldsby *et al.* 2002; Albertson *et al.* 2009; Kennedy *et al.* 2012). In humans, one recent study showed that individuals with inherited defects in DNA polymerases have more somatic mutations accumulating with age and more cancer, yet do not age faster (Robinson *et al.* 2021). Another related study also reported that individuals carrying *MUTYH* germline mutations, a gene involving in DNA repair, have increased somatic mutation rates and were predisposed to colorectal cancer. However, these individuals do not show signs of premature ageing (Robinson *et al.* 2022). Taken together, these studies argue that elevated mutation accumulation alone is unlikely to cause most human ageing-related phenotypes. In addition, while SCNAs and SVs may have a more profound impact than SNVs, they were rarely detected.

**Concluding remarks**

The fact that mutation burden with age is relatively modest, and most somatic mutations are neutral, makes it difficult to picture how somatic mutation accumulation can be a primary driver of ageing. It also means that there is a stark contrast between cancer and ageing: while cancer can originate from mutations in a single cell and subsequent clonal expansion, shown empirically to occur (Evans & DeGregori 2021), age-related dysfunction would need, we suggest, many mutations in a very large number of cells in a tissue. Evolutionarily this has led others to suggest that evolutionary pressure to prevent cancer, a disease widespread across metazoan that can even affect young organisms (Albuquerque *et al.* 2018), will result in levels of somatic mutations across the lifespan that will be far less than are needed to cause most other age-related conditions (de Grey 2007). In other words, if the mutation burden of a lethal conditions like cancer is much lower than that of other ageing phenotypes then, in order to postpone cancer, organisms will evolve to have a mutation burden than is much lower than the number of mutations needed to trigger ageing phenotypes. In this sense, the lower mutation burden observed in long-lived species (Cagan *et al.* 2022) can perhaps be explained by the need to reduce cancer incidence in young animals.

Taken together, recent DNA sequencing experiments focused on quantifying mutations with age reveal a gradual increase in mutations, and widespread evidence of clonal expansion of rapidly dividing mutant clones. These observations are consistent with the age-related increase in cancer observed in most tissues. However, the levels of mutations are relatively small, and difficult to reconcile with most ageing phenotypes. Whether and how somatic mutations in ageing tissues, affecting mostly non-coding regions and overwhelmingly different genes in different cells, can cause dysfunction is unclear. Likewise, while it is possible that clonal expansion is a factor in ageing and results in dysfunction, so far this is not directly supported by experimental data. Recent evidence from inherited mutations in patients with increased somatic mutation burden and no symptoms of accelerated ageing also cast doubt on the role of somatic mutations in most ageing phenotypes. It is possible that other forms of DNA damage and/or genome instability accumulate at much greater rates in human tissues, but these has not been studied in detail and have thus far limited empirical support. The impact of clonal expansion, SCNAs, and SVs on ageing phenotypes, in fact, remain to be further investigated. Advances in genome sequencing technology together with the development of computational methods to reliably detect large-scale structural alterations at a single-cell level should shed light on the potential role of SVs and SCNAs on human ageing.

**Acknowledgements**
K.C. was supported by a Mahidol-Liverpool Ph.D. scholarship from Mahidol University, Thailand, and the University of Liverpool, UK. J.P.M. is supported by grants from the Wellcome Trust (208375/Z/17/Z), Longevity Impetus Grants, LongeCity and the Biotechnology and Biological Sciences Research Council (BB/R014949/1 and BB/V010123/1). We are also grateful to current and past members of our lab for valuable discussions and suggestions, in particular Blaine van Rensburg and Thomas Duffield.